# Low-Temperature Nanoscale Heat Transport in a Gadolinium Iron Garnet Heterostructure Probed by Ultrafast X-ray Diffraction

Deepankar Sri Gyan[1], Danny Mannix[2,3,4], Dina Carbone[5], James L. Sumpter[1], Stephan Geprägs[6], Maxim Dietlein[6,7], Rudolf Gross[6,7,8], Andrius Jurgilaitis[5], Van-Thai Pham[5], Hélène Coudert-Alteirac[5,9], Jörgen Larsson[10,5], Daniel Haskel[11], Jörg Strempfer[11], and Paul G. Evans[1]*

1. University of Wisconsin-Madison, Madison, WI 53706, USA

2. Université Grenoble Alpes, CNRS, Institut Néel, 38042 Grenoble, France

3. European Spallation Source, SE-221 00 Lund, Sweden

4. Aarhus University, Langelandsgade 140, DK-8000 Aarhus, Denmark

5. MAX IV Laboratory, Lund University, P.O. Box 118, SE-221 00 Lund, Sweden

6. Walther-Meißner-Institut, Bayerische Akademie der Wissenschaften, 85748 Garching, Germany

7. Physik-Department, Technische Universität München, 85748 Garching, Germany.

8. Munich Center for Quantum Science and Technology (MCQST), Schellingstraße 7, 80799 München, Germany

9. Department of Physics, University of Gothenburg, SE-41296, Gothenburg, Sweden

10. Department of Physics, Lund University, P.O. Box 118, SE-221 00 Lund, Sweden

11. Advanced Photon Source, Argonne National Laboratory, Lemont, Illinois 60439, USA

*pgevans@wisc.edu

***ABSTRACT***

Time-resolved x-ray diffraction has been used to measure the low-temperature thermal transport properties of a $Pt/Gd_3Fe_5O_{12}//Gd_3Ga_5O_{12}$ metal/oxide heterostructure relevant to applications in spin caloritronics. A pulsed femtosecond optical signal produces a rapid temperature rise in the Pt layer, followed by heat transport into the $Gd_3Fe_5O_{12}$ (GdIG) thin film and the $Gd_3Ga_5O_{12}$ (GGG) substrate. The time dependence of x-ray diffraction from the GdIG layer was tracked using an accelerator-based femtosecond x-ray source. The ultrafast diffraction measurements probed the intensity of the GdIG (1 -1 2) x-ray reflection in a grazing-incidence x-ray geometry. The comparison of the variation of the diffracted x-ray intensity with a model including heat transport and the temperature dependence of the GdIG lattice parameter allows the thermal conductance of the Pt/GdIG and GdIG//GGG interfaces to be determined. Complementary synchrotron x-ray diffraction studies of the low-temperature thermal expansion properties of the GdIG layer provide a precise calibration of the temperature dependence of the GdIG lattice parameter. The interfacial thermal conductance of the Pt/GdIG and GdIG//GGG interfaces determined from the time-resolved diffraction study are of the same order of magnitude as previous reports for metal/oxide and epitaxial dielectric interfaces. The thermal parameters of the Pt/GdIG//GGG heterostructure will aid in the design and implementation of thermal transport devices and nanostructures.

## I.   INTRODUCTION

Thermal transport at the interfaces within magnetic thin film heterostructures depends on the mismatch of phonon dispersions, the conversion between electronic and vibrational heat transport mechanisms, and the overall dimensions of the nanoscale layers.[1,2] In addition to electronic and vibrational effects, heat transport in magnetic insulators can also involve contributions arising from magnetic excitations.[3] These thermal properties have important roles in thermoelectric applications based on the spin Seebeck effect (SSE) and in heat-assisted magnetic recording (HAMR).[1,4-6] Understanding thermal transport in nanoscale thin film heterostructures is a challenge because the structural, vibrational, and electronic as well as magnetic properties of nanoscale layers and their interfaces can differ significantly from bulk materials.[7,8] In particular, the interfacial thermal conductance is increasingly important and challenging to measure in nanoscale layers. Thermal transport is described in heat diffusion models using parameters such as the thermal conductivity and heat capacity of each layer and the thermal conductance of the interfaces between them.[9] Theoretical predictions of interfacial thermal conductance describe the conversion of a thermal current carried primarily by electrons in the metal and to phonons in the insulator and the effects of compositional grading, intermixing, and roughness.[1,10-14] Effects arising from uncertainties in the structure of the interfaces and the contributions from magnon-phonon interactions make accurate theoretical predictions of the interfacial thermal conductance challenging.

Experimental measurements complement these predictions and provide the means for the design of devices incorporating thermal transport effects. Thermal transport can be observed in time-resolved x-ray diffraction experiments by interpreting structural changes induced by transient heating. The experimental timescale of thermal transport in nanoscale layers following a heat impulse can be estimated using the thermal diffusivity and the layer thickness. For a GdIG layer with a thickness of 100 nm, this time is on the order of tens to hundreds of ps, based on the thermal parameters described below. The timescale of photoacoustic propagation can be estimated using the layer thickness and the longitudinal acoustic sound velocity.[15,16] The longitudinal acoustic sound velocities in GdIG and GGG are 6.50 km s$^{-1}$ and 6.34 km s$^{-1}$, respectively.[17] Acoustic propagation is thus complete within approximately 15 ps, far shorter than the thermal evolution over the 100 ps that is apparent in the experiments reported below. Previous studies have demonstrated that it is possible to determine the thermal properties of nanoscale layers by comparing the measured x-ray diffraction patterns with the models of thermal transport. Examples include the thermal conductivity of semiconductor and complex oxide thin film heterostructures and graphite thin films.[18-20] Here, the changes in the diffracted x-ray intensity as a function of time are monitored to track thermal transport in a magnetic insulator thin film in the temperature regime relevant to SSE devices. A thermal transport model was used to determine the interfacial thermal conductances in the heterostructure. Time-resolved x-ray measurements are particularly useful in probing heat transport at these nanometer length scales because the transient heating leads to large temperature gradients that make it possible to study the interface conductance.

The optical-pump x-ray-probe experiment studied nanoscale low-temperature thermal transport in a Pt/Gd$_3$Fe$_5$O$_{12}$ (GdIG)//Gd$_3$Ga$_5$O$_{12}$ (GGG) heterostructure. The Pt layer was heated by an ultrafast optical pulse, producing a pulse of heat that was transported through the Pt/GdIG interface and subsequently into the GGG substrate. The data-acquisition time step employed in the x-ray diffraction experiment was 4 ps, which was sufficient to probe the timescale at which thermal transport effects perturb the structure. The x-ray and optical pulse durations were shorter than this time step, as described below. The thermal transport processes that follow optical excitation are illustrated in Fig. 1(a). The time interval between the optical pulse and the x-ray probe pulse is $t$. Figure 1(b) shows the temperature profile for $t$=0 arising from the absorption of optical energy in the metallic Pt layer, immediately after a $\pi$-polarized optical pulse with a wavelength of 395 nm and a fluence of 4.1 mJ cm$^{-2}$ for an initial sample temperature $T_{initial}$=16 K. A detailed thermal transport calculation, for which the $t$=0 result is shown in Fig. 1(b), is described below. The temperature profile in Fig. 1(b) represents the initial condition for models of the heat transport through the Pt/GdIG//GGG heterostructure. The thermal conductances of the Pt/GdIG and GdIG//GGG interfaces were determined by modeling the thermal transport through the heterostructure and fitting the time dependence of the intensity of the diffracted x-ray beam.

The interfacial conductance for metal/oxide interfaces is in general nearly independent of temperature near room temperature and exhibits linear temperature dependence at low temperature.[21] The lattice parameter, specific heat, and thermal conductivity of bulk rare-earth garnets vary strongly as a function of temperature and can exhibit significant temperature dependence of the thermal expansion behavior in some cases. For example, a negative thermal expansion coefficient is observed in $Tb_3Ga_5O_{12}$ at temperatures below 50 K.[22] The low-temperature thermal expansion behavior of GdIG thin films epitaxially grown on GGG substrate can thus be expected to depend strongly on temperature. The temperature dependence of the GdIG lattice parameter is important in the interpretation of the time-resolved x-ray scattering results. This work thus includes a study of low-temperature thermal expansion phenomena in the GdIG thin film layer.

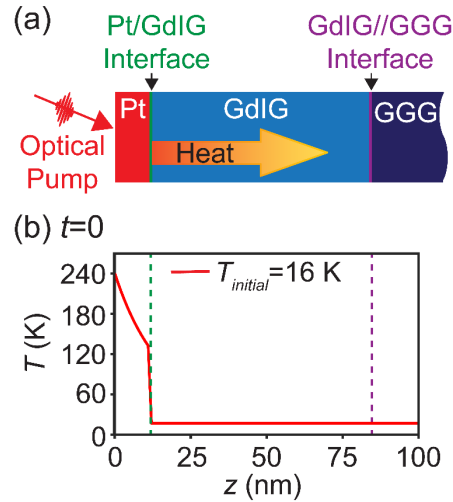

FIG. 1. (a) Heat transport following optical absorption in the Pt layer of a Pt/GdIG//GGG heterostructure. (b) Temperature profile at $t$=0, immediately following optical excitation under the experimental conditions for an initial temperature ($T_{initial}$) of 16 K. The surface of the heterostructure is at depth $z$=0.

## II.    EXPERIMENT

### A.    Time-resolved x-ray diffraction

Time-resolved x-ray diffraction experiments were conducted at the FemtoMAX facility at the MAX IV Laboratory.[23] The Pt/GdIG//GGG heterostructure consisted of a 13 nm Pt layer deposited on a 75-nm-thick (111)-oriented GdIG film on a (111) GGG substrate.[24] Details of the preparation of the GdIG and Pt layers are provided in refs. [24] and [30]. The x-ray probe pulses had a photon energy of 7.880 keV, 2 Hz repetition rate, and a pulse duration less than 200 fs. The time-resolved diffraction experiment employed a $z$-axis x-ray diffractometer with the probe x-ray beam incident in grazing-incidence geometry at a grazing incidence angle $\alpha$=0.5°.[25] The optical pump employed the parameters given above and had an angle of 20° with respect to the x-ray probe. The effective time resolution of the experiment was 0.8 ps due to the non-collinearity of the optical pump and x-ray probe pulses. In this grazing-incidence x-ray scattering geometry, the intensity of the incident x-ray beam reached $1/e^2$ of the incident intensity at a depth of 12 nm. The relatively short x-ray attenuation depth in this geometry reduces the intensity of the substrate x-ray reflection. The x-ray exit angle is much greater than the incident angle, so the probed volume was determined by the incident beam penetration. The optical absorption length for Pt at a wavelength of 395 nm is $\zeta$=10 nm.[26] The majority of the optical pump power is absorbed in the Pt layer. The optical absorption lengths at 395 nm for GdIG and GGG are 360 nm and 4 cm, respectively, both far longer than in Pt.[27,28] The heating of GdIG and GGG layers is thus a consequence of thermal transport from the heated Pt layer, rather than optical absorption. The x-ray probe spot size on the sample under the experimental conditions was 23 mm × 0.27 mm, smaller than the optical spot size of 28 mm × 1.36 mm.

The experimental arrangement and a Cartesian coordinate system defining the diffraction geometry are shown in Fig. 2(a). The diffraction angle $2\theta_h$ is defined as the angle between the projection of the diffracted beam on the $x$-$y$ plane with the $y$-axis. $2\theta_v$ is the angle between the diffracted x-ray beam and the $x$-$y$ plane.[29] The azimuthal angle of the incident beam with respect to the $y$ axis is $\varphi$. The diffracted beam was collected using a charge-coupled device (CCD) x-ray detector (IKON, Andor, Inc.).

The intensity of the GdIG (1 -1 2) reflection was recorded as a function of pump-probe delay at fixed diffraction angles of $2\theta_h$=15.32° and $2\theta_v$=8.14°. Diffraction patterns were recorded with a delay step of 4 ps for times up to 600 ps following the optical excitation. The experiment was performed at two values of $T_{initial}$: 16 K and 77 K. The temperature range was selected based on spin-Seebeck measurements on the Pt/GdIG//GGG devices. The spin-Seebeck signal shows a sign change at 287 K, the magnetic compensation temperature of GdIG.[30] An additional sign change was observed at near 70 K which is attributed to competing magnon modes in GdIG.[30] The measurements at a lower temperature of $T_{initial}$=16 K and at a temperature slightly above 70 K, i.e., $T_{initial}$=77 K approximately match the temperature regime of the low-temperature sign change.

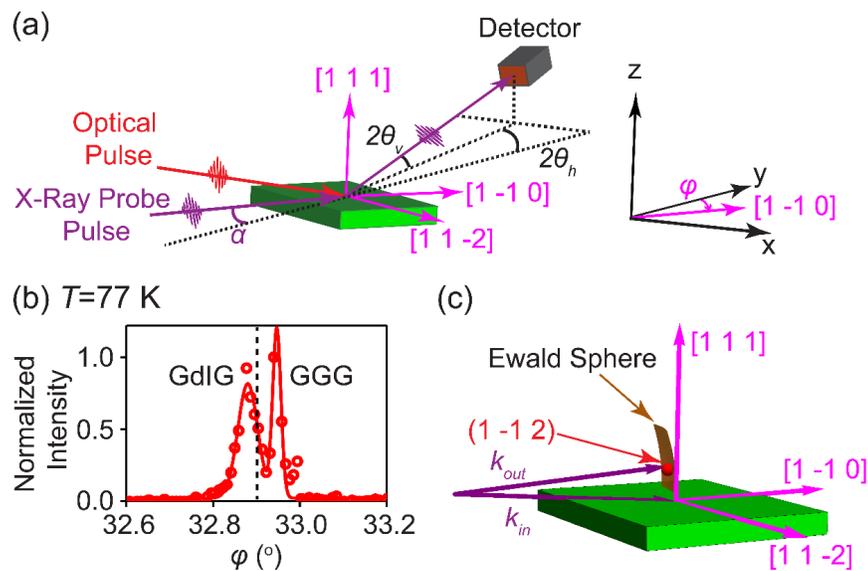

FIG. 2. (a) Experimental arrangement for the time-resolved x-ray diffraction experiment. (b) Measured (points) and fit (line) for a rocking curve scan as a function of $\varphi$ at $T_{initial}$=77 K. The dashed line shows the angular setting at which the time-resolved diffraction study was conducted. (c) Reciprocal space for the GdIG layer showing the (1 -1 2) reflection and the Ewald sphere for the scattering geometry of the time-resolved diffraction measurement. $k_{in}$ and $k_{out}$ represent the incident and diffracted x-ray wavevectors.

A rocking curve scan of the intensity of the (1 -1 2) reflection as a function of $\varphi$ is shown in Fig. 2(b). The two intensity maxima in Fig. 2(b) correspond to the (1 -1 2) x-ray reflections of the GdIG layer and the GGG substrate, respectively. At $T_{initial}$=77 K the GdIG layer has an intensity maximum at $\varphi_{(1\text{ -}1\,2)}$=32.87°. The GGG (1 -1 2) reflection appears at a higher value of $\varphi$ than the GdIG layer. A diagram of reciprocal space for the GdIG layer is shown in Fig. 2(c), including the Ewald sphere and the (1 -1 2) reflection of GdIG. The GdIG (1 -1 2) diffracted intensity profile in Fig. 2(b) was fitted with a Gaussian intensity distribution to determine the parameters used in the analysis below. The intensity model is:

$$I(\varphi) = y_o + I_{max} \exp\left(-2\left(\frac{\varphi - \varphi_{(1\,-1\,2)}}{w}\right)^2\right) \qquad (1)$$

Here, $w$=0.0462° is the angular width of the GdIG reflection in the $\varphi$ angle, $I(\varphi)$ is the intensity of the film reflection normalized to the intensity of the substrate reflection, $I_{max}$=0.79 is the peak intensity of the GdIG thin film reflection normalized to the substrate, and $y_o$=0.018 is the background. The time-resolved x-ray diffraction patterns below were recorded with $\varphi=\varphi_0$=32.90°. In an ideal case, the measurement of the diffracted intensity consists of tracking the full rocking curve at each time step. These measurements were not pursued, however, because of the relatively long acquisition time that would have been required. The signal-to-noise ratio of the fixed-incident-angle measurements was sufficient to observe the intensity change corresponding to nanoscale thermal transport.

Diffraction patterns of the GdIG (1 -1 2) reflection at $T_{initial}$=16 K before the optical excitation ($t$<0) and at a delay time of 200 and 500 ps following the optical excitation are shown in Fig. 3(a). The centers of mass of the intensity in the diffraction patterns in $2\theta_h$ and $2\theta_v$ are shown as a function of delay time in Fig. 3(b). The temperature increase in the GdIG thin film under the experimental conditions for $T_{initial}$=16 K changes the lattice parameter of the GdIG. A possible artifact would arise if the diffraction measurement tracked the intensity of the Bragg peak rather than sampling the intensity at a fixed wavevector along the GdIG truncation rod. This artifact would arise, for example in the case when the incident-beam angular divergence is large. If the diffraction measurement tracked the intensity of the Bragg peak the expected shifts in $2\theta_h$ and $2\theta_v$ would be 0.002º and 0.006º, respectively. A shift of this magnitude would be more than the experimental uncertainty in the case of $2\theta_v$ (±0.0012) but is not observed in Fig. 3(b). The absence of an angular shift in the angles at which the GdIG (1 -1 2) x-ray reflection appears in Fig. 3 indicates that the intensity change in the experiments is due to the change in intensity at a fixed wavevector along the truncation rod. The intensity of the diffraction peak can thus be interpreted to extract structural information.

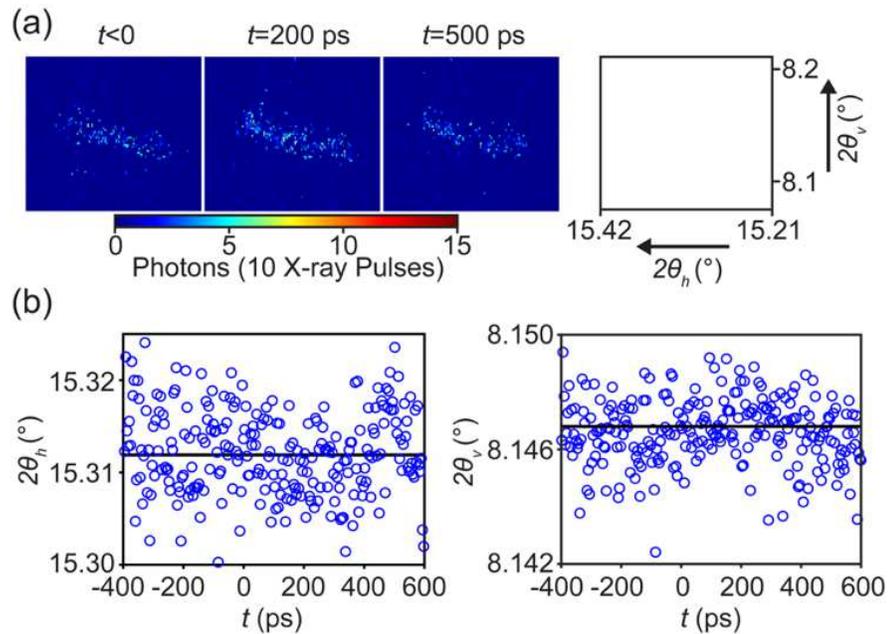

FIG. 3. (a) Diffraction patterns for the GdIG (1 -1 2) reflection for $T_{initial}$=16 K before excitation ($t$<0) and for $t$=200 and 500 ps. The range of $2\theta_h$ and $2\theta_v$ mapped in the detector images is shown at right. (b) Time-dependence of the angular center of mass of the diffracted intensity in $2\theta_h$ and

*2θc.*

**B. Temperature dependence of the lattice parameter of epitaxially constrained GdIG at low temperature**

The temperature dependence of the shift in the out-of-plane wavevector of the GdIG layer was measured at station 4-ID-D of the Advanced Photon Source at Argonne National Laboratory. X-ray diffraction patterns of the GdIG (4 4 4) and GGG (4 4 4) reflections were recorded during heating from 10 K to 150 K in steps of 30 K with a photon energy of 7.9345 keV. The diffraction patterns were acquired as *θ-2θ* scans through the GdIG (4 4 4) and GGG (4 4 4) reflections.

**III.    RESULTS AND DISCUSSION**

**A. Low-temperature steady-state temperature dependent $d_{444}$ of GdIG thin film**

Diffraction patterns for the (4 4 4) reflection of GdIG and GGG were measured under steady-state conditions at temperatures from 10 to 150 K. Diffraction patterns obtained in *θ-2θ* scans at the extremes of this temperature range are shown in Fig. 4(a). The wavevector $Q_{444}$ of the GdIG and GGG (4 4 4) reflections were determined by fitting the diffraction profiles with a Gaussian function. The temperature dependence of the difference between these wavevectors, $Q_{444,GGG}$ - $Q_{444,GdIG}$, is shown in Fig. 4(b). The temperature dependence of the lattice parameter of the GGG substrate has been previously reported.[31] The difference between the wavevectors of the GGG substrate and the GdIG film was used along with the reported GGG lattice parameter to calculate the temperature dependence of $d_{444}$ for the GdIG film shown in Fig. 4(c). For GdIG, $d_{444}$ decreases with temperature up to 90 K and then increases throughout the remainder of the experimental temperature range. There is a lattice contraction between 10 and 90 K and a lattice expansion with a much smaller magnitude at higher temperatures. The temperature dependence of the lattice parameter of the GdIG layer is qualitatively similar to previous observations for $Tb_3Ga_5O_{12}$, including a low-temperature contraction and expansion at higher temperatures.[22]

The temperature dependence of $d_{444}$ for GdIG shown in Fig. 4(c) corresponds to steady-state conditions in which the GGG substrate and GdIG thin film are at the same temperature. Pulsed optical excitation, however, heats the GdIG film under the conditions in which the GGG substrate remains at $T_{initial}$. The interpretation of the time-resolved study depends on the lattice parameter of the GdIG layer under conditions in which there is an additional constraint applied by maintaining the substrate at a constant temperature. Figure 4(d) shows the measured $d_{444}$ of the GdIG layer as a function of the GdIG temperature corrected for the conditions in which the substrate remains at a fixed temperature, i.e., $T_{initial}$=16 K and $T_{initial}$=77 K. The elastic correction was made based on the previously reported thermal expansion measurement of GGG.[31] For $T_{initial}$=16 K, the GdIG thin film under these elastic conditions undergoes a lattice contraction as the temperature increases to 90 K and then expands for temperatures above 90 K. For $T_{initial}$=77 K, the GdIG thin film expands at all higher temperatures.

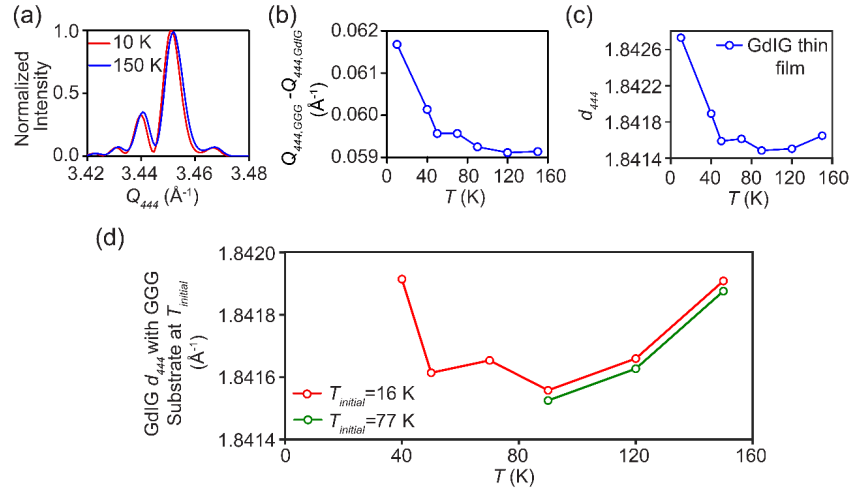

FIG. 4. (a) Diffraction patterns of the GdIG (4 4 4) reflection at 10 K and 150 K. (b) Spacing between measured wavevectors of the GdIG (4 4 4) and GGG (4 4 4) reflections, as a function of temperature under the condition when both GGG and GdIG are at the same temperature. (c) $d_{444}$ for the GdIG thin film as a function of temperature based on the measured wavevector difference in (b) and the previously reported temperature dependence of the lattice parameter of GGG.[31] (d) $d_{444}$ for the GdIG thin film as a function of temperature, corrected for the elastic conditions that apply when the GdIG layer is heated and the GGG substrate is held at $T_{initial}$.

## B. Time-resolved x-ray diffraction intensity measurements

The intensity of the (1 -1 2) reflection was measured as a function of pump-probe delay time at a fixed azimuthal angle $\varphi_0 = 32.90°$. This angle corresponds to a setting between the GdIG and GGG reflections in Fig. 2(b). Heating can be expected to shift the GdIG (1 -1 2) reflection to a higher or a lower angle, due to the contraction or expansion of the crystal. Accordingly, the diffracted intensity measured at $\varphi_0$ increases or decreases upon heating and can be used as a marker for the peak shift. The observed time dependences of the diffracted intensities for $T_{initial}$=16 K and $T_{initial}$=77 K are shown in Figs. 5(a) and (b), respectively. The diffraction patterns were acquired with 10 x-ray pulses at each time step at an x-ray pulse rate of 2 Hz. The intensity was integrated over the diffraction peak, with a smooth background due to diffuse scattering subtracted, and averaged for all 10 pulses to determine the intensity used in the analysis. For $T_{initial}$=16 K, Fig. 5(a), the intensity of the diffracted beam shows a 20% increase for $t$ up to 200 ps. The intensity then relaxes towards the initial value for the remaining duration of the measurement, up to 600 ps. The uncertainty in the integrated intensity arising from shot noise was 4.5% for measurements with $T_{initial}$=16 K. The observed intensity change is much higher than the uncertainty caused by shot noise. The intensity of the diffracted beam for $T_{initial}$=77 K exhibits changes that are smaller than the experimental uncertainty, as shown in Fig. 5(b). The diffracted intensity was lower for the $T_{initial}$=77 K measurement, leading to a 10% uncertainty. The origin of the difference between the intensity profiles for the two initial temperatures is discussed below. Figure 5 also shows the result of simulations of the intensity, which are discussed in section C.

The possibility that the Debye-Waller effect contributes to the observed change in intensity can be evaluated using the steady-state temperature dependent diffraction measurements. The observed decrease in the integrated intensity of the GdIG (4 4 4) reflection between 20 K and 50 K in the steady-state experiments was 2%. Even at this higher wavevector, the intensity decrease is much smaller than the 20% increase observed in the optically excited measurements with the (1 -1 2) reflection. Intensity changes due to the Debye-Waller effect were thus neglected in the analysis.

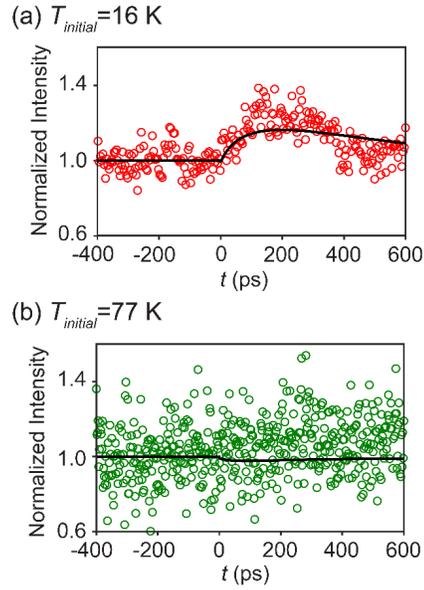

(a) $T_{initial}$=16 K

(b) $T_{initial}$=77 K

FIG. 5. Time dependence of the observed (points) and simulated (lines) diffracted x-ray intensity at $\varphi = \varphi_0$ after optical excitation at $t$=0 for (a) $T_{initial}$=16 K, and (b) $T_{initial}$=77 K.

**C. Simulation of optically induced heating and thermal transport**

The absorbed optical energy per unit volume in the Pt layer is:[15,22]

$$W(z) = (1 - R)\frac{F}{A\zeta}e^{-\frac{z}{\zeta}} \qquad (2)$$

Here $R$=0.4 is the optical reflectivity of Pt at a wavelength of 395 nm for the π-polarized optical pump at the experimental incident angle and $\zeta$ is the optical absorption length given above.[32] $F$ is the incident optical energy per pulse, $A$ is the illuminated area, and $z$ is the depth with the surface at $z$=0. The optical absorption in the Pt/GdIG//GGG heterostructure was also calculated more precisely by considering the complex effects arising from refraction and interference.[33] The calculation using this more complete approach produces a similar temperature profile within the GdIG layer as the Lambert-Beer approximation. The analysis has thus, for simplicity, been based on the approach shown in Eq. 2.

The heat required to raise the temperature from $T_{initial}$ to $T(z,t=0)$ is:

$$W(z) = \int_{T_{initial}}^{T(z,t=0)} C(z,T)dT \qquad (3)$$

Here $C(z,T)$ is the heat capacity per unit volume, the product of specific heat capacity $c_p$, and the mass density $\rho$. $T(z,t=0)$ was determined by solving Eqs. 2 and 3. The calculated $T(z,t=0)$ for $T_{initial}$=16 K is shown in Fig. 1(b).

A thermal diffusion equation was solved to obtain the temperature distribution in the Pt/GdIG//GGG heterostructure as a function of depth and time:[9]

$$\rho c_p \frac{\partial T(z,t)}{\partial t} = \kappa \frac{\partial^2 T(z,t)}{\partial z^2} \qquad (4)$$

The values of the thermal conductivity $\kappa$ and $c_p$ are different for the Pt, GdIG, and GGG layers. The temperature before optical excitation, $T_{initial}$, was used as the boundary condition at large substrate depth. The simulation considered a depth of 1.2 μm. The thermal transport properties of the bulk forms of the components of the heterostructure and their interfaces are highly temperature dependent in the temperature range of this experiment.[34,35] The thermal conductivities of garnets such as $Y_3Fe_5O_{12}$ (YIG) and GGG show a common trend at low temperatures, reaching a maximum at approximately 30 K.[35,36] The calculation presented here uses the approximation that the thermal conductivity of the GdIG layer has a value and temperature dependence equal to YIG and that the thermal conductivity and heat capacity of GGG have their bulk values.[36-38] The simulations use the bulk heat capacity of GdIG.[39] The specific heat of GdIG has been reported only in the temperature range up to 30 K but it shows similar variation in that range to GGG.[39] The temperature dependences of the specific heat for GdIG and GGG were assumed to be the same and were used to calculate the specific heat of GdIG. The calculation uses reported values of the thermal conductivity and heat capacity of Pt.[34,40,41] The values of the parameters used for the thermal diffusion simulation at 16 K and 77 K are summarized in Table 1.

Table 1: Values of specific heat and thermal conductivity of the components of the Pt/GdIG//GGG heterostructure used for simulations at $T_{initial}$=16 K and $T_{initial}$=77 K. Values at 16 K and 77 K are interpolated from the temperature dependence reported in the references.

| Layer | Parameter | Value at 16 K * | Value at 77 K * | Reference |
|-------|-----------|-----------------|-----------------|-----------|
| Pt | Specific heat ($c_p$) | 0.0031 J g$^{-1}$ K$^{-1}$ | 2.11 J cm$^{-3}$ K$^{-1}$ | [40] [41] |
| | Thermal conductivity ($\kappa$) | 741.39 W m$^{-1}$ K$^{-1}$ | 86.29 W m$^{-1}$ K$^{-1}$ | [34] |
| GdIG | Specific heat ($c_p$) | 0.0325 J cm$^{-3}$ K$^{-1}$ | 1.01 J cm$^{-3}$ K$^{-1}$ | [38] [39] |
| | Thermal conductivity ($\kappa$) | 90.27 W m$^{-1}$ K$^{-1}$ | 34.92 W m$^{-1}$ K$^{-1}$ | [36] |
| GGG | Specific heat ($c_p$) | 0.0325 J g$^{-1}$ K$^{-1}$ | 1.01 J cm$^{-3}$ K$^{-1}$ | [38] |
| | Thermal conductivity ($\kappa$) | 530.65 W m$^{-1}$ K$^{-1}$ | 44.25 W m$^{-1}$ K$^{-1}$ | [37] |

The surface is assumed to be insulated so that $\partial T/\partial z$=0 at $z$=0. There are also four interfacial boundary conditions. The boundary conditions at the Pt/GdIG interface are:

$$-\kappa_{pt}\frac{\partial T_{Pt}}{\partial z} = h_{Pt/GdIG}(T_{Pt} - T_{GdIG}) \qquad \text{for } z=d_{\text{GdIG}}{}^-$$

$$-\kappa_{GdIG}\frac{\partial T_{GdIG}}{\partial z} = h_{Pt/GdIG}(T_{Pt} - T_{GdIG}) \qquad \text{for } z=d_{\text{GdIG}}{}^+$$

The boundary conditions at the GdIG//GGG interface are:

$$-\kappa_{GdIG}\frac{\partial T_{GdIG}}{\partial z} = h_{GdIG//GGG}(T_{GdIG} - T_{GGG}) \quad \text{for } z=d_{\text{GGG}}{}^-$$

$$-\kappa_{GGG}\frac{\partial T_{GGG}}{\partial z} = h_{GdIG//GGG}(T_{GdIG} - T_{GGG}) \quad \text{for } z=d_{\text{GGG}}{}^+$$

Here $h_{Pt/GdIG}$ and $h_{GdIG//GGG}$ are the interfacial thermal conductances of the Pt/GdIG and GdIG//GGG interfaces, respectively, and $d_{GdIG}$ and $d_{GGG}$ are the depths of Pt/GdIG and GdIG//GGG interfaces with respect to the surface. The superscript notations + and – following $d_{GdIG}$ and $d_{GGG}$ indicate the low-$z$ and high-$z$ sides of these interfaces, respectively. The values of interfacial thermal conductance were assumed to be proportional to temperature, as discussed in section D.

The thermal diffusion equation was solved using the forward Euler method.[42] Figures 6(a) and (b) show the simulated temperature distribution in the Pt/GdIG//GGG heterostructure as a function of $z$ and $t$. These simulations employ the interface conductances determined using the process described in section D. A comparison of the simulation results in Figs. 6(a) and (b) show that the Pt layer stays at a higher temperature for a longer time for $T_{initial}$=16 K than for $T_{initial}$ =77 K.

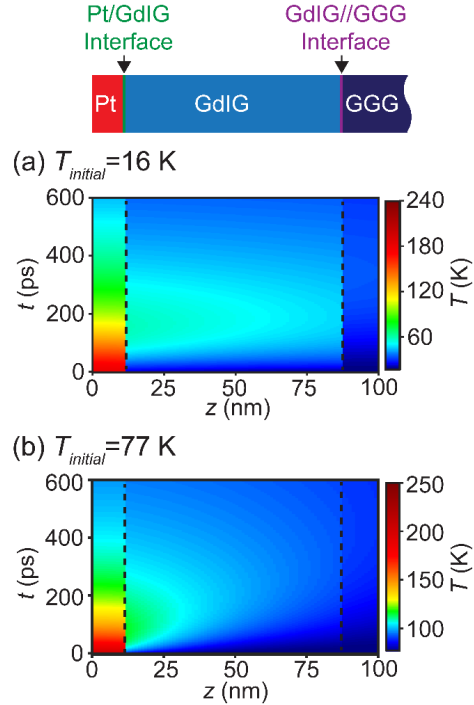

FIG. 6. (a) Simulated variation of temperature with time and depth in the Pt/GdIG//GGG heterostructure for $T_{initial}$=16 K and (b) $T_{initial}$=77 K.

Figure 7(a) shows the predicted time dependence of $\Delta T_{avg}$, the depth-averaged increase in the temperature of the experimentally probed volume of the GdIG layer. The depth average was computed as a volume average and was not weighted by the depth dependence of the incident intensity. The maximum reached by $\Delta T_{avg}$ is larger for simulations at $T_{initial}$=16 K than for simulations at $T_{initial}$=77 K because of the larger heat capacity of GdIG at $T_{initial}$=77 K. The temperature in the GdIG layer for $T_{initial}$=16 K increases for $t$ up to 200 ps and then relaxes towards $T_{initial}$. The measured diffracted intensity for $T_{initial}$=16 K, shown in Fig. 5(a), has a time dependence that is similar to the time dependence of the simulated temperature. Both the measured intensity and the simulated temperature increase reach a maximum at approximately 200 ps. The $\Delta T_{avg}$ profile for $T_{initial}$=77 K also increases for up to 200 ps and shows a slower relaxation than for $T_{initial}$=16 K.

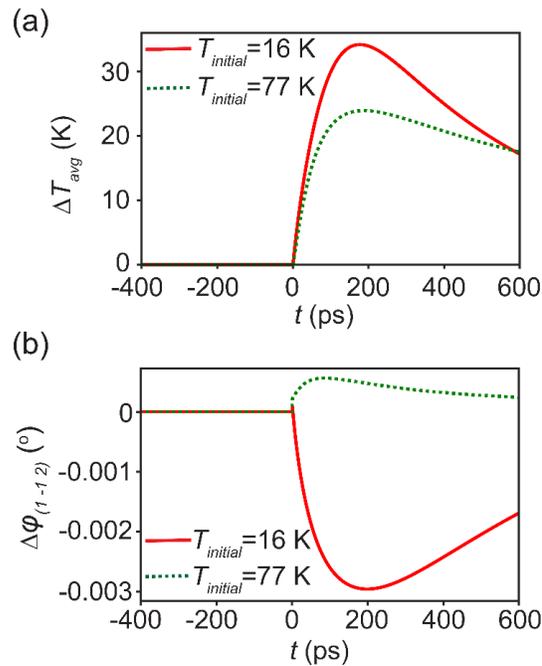

FIG. 7. (a) Simulated depth-averaged temperature increase in the volume of the GdIG layer probed by the time-resolved x-ray diffraction experiment as a function of $t$ for $T_{initial}$=16 K and $T_{initial}$=77 K. (b) Predicted angular shift $\Delta\varphi_{(1\text{ -}1\text{ }2)}$ of the (1 -1 2) reflection corresponding to the temperature increase in the GdIG layer for $T_{initial}$=16 K and $T_{initial}$=77 K.

The azimuthal angle $\varphi_{(1\text{ -}1\text{ }2)}$ of the peak intensity of the GdIG (1 -1 2) reflection was calculated as a function of temperature using the calculated temperature and the measured temperature-dependent lattice parameter of GdIG. The lattice expansion or contraction caused by the temperature rise in the GdIG layer causes a shift $\Delta\varphi_{(1\text{ -}1\text{ }2)}$ in the peak of the GdIG (1 -1 2) reflection with respect to its initial value. In order to compare the thermal simulation and the diffraction results, the simulated temperature in the top 12 nm of the GdIG was used to calculate the lattice parameter as a function of depth. The value of $\Delta\varphi_{(1 \text{ -}1\ 2)}$ was calculated at each time using a kinematic diffraction calculation based on these lattice parameters. The variation of $\Delta\varphi_{(1 \text{ -}1\ 2)}$ with time was calculated using the temperature distribution in the GdIG layer shown in the temperature maps in Figs. 6(a) and (b). The variation of $\Delta\varphi_{(1 \text{ -}1\ 2)}$ with time for $T_{initial}$=16 K and $T_{initial}$=77 K is shown in Fig. 7(b).

We assume that the diffracted intensity at each time step was described using the Gaussian function, but with a distribution $\Gamma'(\varphi)$ determined by a different value of $\varphi_{(1 \text{ -}1\ 2)}$ corresponding to the elevated temperature of the GdIG thin film. The maximum of the distribution $\Gamma'(\varphi)$ is at $\varphi_{(1 \text{ -}1\ 2)}+\Delta\varphi_{(1 \text{ -}1\ 2)}$. The simulated time dependence of the diffracted intensity was found by calculating $\Gamma'(\varphi_0)$ at each time step. The intensities were normalized to the value corresponding to $t<0$. The simulated variations of normalized intensity with time for $T_{initial}$=16 K and $T_{initial}$=77 K are shown in Figs. 5(a) and (b).

For $T_{initial}$ =16 K, the maximum value of $\Delta T_{avg}$ was 35 K, reached at $t$=200 ps. The  and the maximum temperature in the GdIG layer was 56 K. Both of these maximum temperatures indicate that at all values of $t$ the GdIG layer remained in the temperature regime of thermal contraction for $T_{initial}$ =16 K. In this regime, the GdIG (1 -1 2) reflection shifts to higher $\varphi$ upon heating, leading to an increase in the intensity at the angular setting of the time-resolved measurement, as observed in the experiment. The simulated intensity variation at $T_{initial}$=77 K exhibits an intensity change that is smaller than the experimental uncertainty. The difference between $T_{initial}$=16 K and $T_{initial}$=77 K can be interpreted as arising from the difference in the thermal expansion for these two initial temperatures. $T_{initial}$=77 K is close to the temperature where the film undergoes a transition from a thermal contraction to a regime of thermal expansion. The coefficient of thermal expansion of the GdIG layer reaches a minimum near 77 K, leading to a significantly lower change in intensity for $T_{initial}$=77 K than for $T_{initial}$=16 K.

**D. Determination of interfacial thermal conductance**

The key unknown parameters in the thermal transport from the Pt layer to the GdIG and GGG layers are the interfacial thermal conductances $h_{Pt/GdIG}$ and $h_{GdIG/GGG}$. The thermal conductances of metal/dielectric or dielectric/dielectric interfaces are highly temperature-dependent at low temperatures and depend on factors such as the nature of interfacial roughness and the acoustic mismatch at the interface.[21,43,44] A previous comparison of experimental data with theoretical calculations suggests that for both metal/dielectric and dielectric/dielectric contacts in the 10-200 K temperature range a linear variation of thermal conductance with temperature is a reasonable approximation.[21] The model described here thus assumes that the thermal conductances of the Pt/GdIG and GdIG//GGG interfaces are proportional to temperature in the range of temperatures in this measurement. The interfacial thermal conductances are assumed to be:

$$h_{Pt/GdIG} = m_{Pt/GdIG}T$$

and $h_{GdIG//GGG} = m_{GdIG//GGG}T$

Here $m_{Pt/GdIG}$ and $m_{GdIG//GGG}$ are constants.

The diffracted intensity at fixed incident angle $\varphi_0$ was simulated for a range of values of $m_{Pt/GdIG}$ and $m_{GdIG//GGG}$ and compared to the measured intensity. The x-ray intensity was simulated using a temperature profile based on the thermal parameters in Section C. The shift in the azimuthal angle of the (1 -1 2) reflection was calculated using the measured thermal variation of the lattice parameter and used to determine the intensity at the measurement setting $\varphi_0$, as described above. The simulation was repeated for a range of values of the parameters $m_{Pt/GdIG}$ and $m_{GdIG//GGG}$. The best fit for the intensity variation with time, shown as solid lines in Fig. 5(a) and (b), was observed for $m_{Pt/GdIG}$=0.0035 GW m$^{-2}$ K$^{-2}$ and $m_{GdIG//GGG}$ =0.02 GW m$^{-2}$ K$^{-2}$. Changes in the value of $m_{Pt/GdIG}$ primarily affect the early $t$ rising edge of the intensity variation. Values of $m_{Pt/GdIG}$ in the range 0.003 GW m$^{-2}$ K$^{-2}$ to 0.004 GW m$^{-2}$ K$^{-2}$ fit the rise in intensity well, with lower and higher values correspondingly making the initial rise faster or slower. The simulated diffracted intensity depends strongly on $m_{GdIG//GGG}$ for $t$ greater than 200 ps. In this time regime, values of $m_{GdIG//GGG}$ ranging from 0.016 to 0.024 GW m$^{-2}$ K$^{-2}$ provide a good match to the time dependence of the intensity. Based on this value of $m_{Pt/GdIG}$, the interfacial thermal conductance of the Pt/GdIG interface at 16 K and 77 K are 0.06 GW m$^{-2}$ K$^{-1}$ and 0.27 GW m$^{-2}$ K$^{-1}$, respectively. The interfacial thermal conductance of the GdIG//GGG interface at 16 K and 77 K are 0.34 GW m$^{-2}$ K$^{-1}$ and 1.54 GW m$^{-2}$ K$^{-1}$.

We can compare these values with reports for other metal/dielectric and epitaxial dielectric/dielectric interfaces. The thermal conductance of the epitaxial TiN//MgO (001), TiN//MgO(111), and TiN//Al$_2$O$_3$(0001) interfaces are approximately 0.4 GW m$^{-2}$ K$^{-1}$ at 100 K and 0.2 GW m$^{-2}$ K$^{-1}$ at 80 K, of the same order of magnitude as the GdIG//GGG interface thermal conductance reported here.[44] The thermal conductances of the interfaces between Ti, Al, Au, and Pb metals and Al$_2$O$_3$ ranges from 0.03 to 0.08 GW m$^{-2}$ K$^{-1}$ at 80 K, similar to the Pt/GdIG interface.[45]

**IV. CONCLUSION**

A combination of ultrafast optical excitation and time-resolved x-ray diffraction allows the low-temperature thermal transport properties of the Pt/GdIG and GdIG//GGG interfaces to be determined. The effectively instantaneous temperature rise in the Pt layer following optical excitation leads to a heat pulse that diffuses through the Pt/GdIG interface, the GdIG layer, the GdIG//GGG interface, and into the GGG substrate. Steady-state measurements of the thermal expansion properties of the GdIG layer show that the GdIG layer exhibits thermal contraction below 90 K and thermal expansion at higher temperatures. For an initial temperature of 16 K, time-resolved experiments at a fixed incident x-ray angle exhibit an increase in the diffracted beam intensity for 200 ps following the optical excitation and a return to the initial value at later times. The variation in the diffracted intensity under the same settings is smaller than the experimental uncertainty for an initial temperature of 77 K. Fitting the time-dependence of the intensity profile with the assumptions used in our analysis provides a reasonable estimate of the values of the interfacial thermal conductance of Pt/GdIG and GdIG//GGG. These results will assist in the interpretation of spin-Seebeck effect experiments because the temperature profile in electrical measurements is often estimated using diffusive models depending on the interfacial thermal conductance.[21,30,43]

### ACKNOWLEDGMENT

The authors acknowledge support from the U.S. DOE, Basic Energy Sciences, Materials Sciences and Engineering, under contract no. DE-FG02-04ER46147. The time-resolved x-ray diffraction experiment was performed at the FemtoMAX beamline of MAX IV synchrotron source in Lund, Sweden, funded by the Knut and Alice Wallenberg Foundation and 12 Swedish universities (KAW, No. 2010.0098). J.L. acknowledges support from the Swedish Research Council (VR, Grant No. 2015-06115). The low-temperature thermal expansion measurement was performed at the 4-ID-D beamline of Advanced Photon Source, a U.S. Department of Energy Office of Science User Facility operated for the DOE Office of Science by Argonne National Laboratory under Contract No. DE-AC02-06CH11357.

**AUTHORS DECLARATION**

**Conflict of Interest**

The authors have no conflicts to disclose.

**DATA AVAILABILITY**

The data that support the findings of this study are available from the corresponding author upon reasonable request.

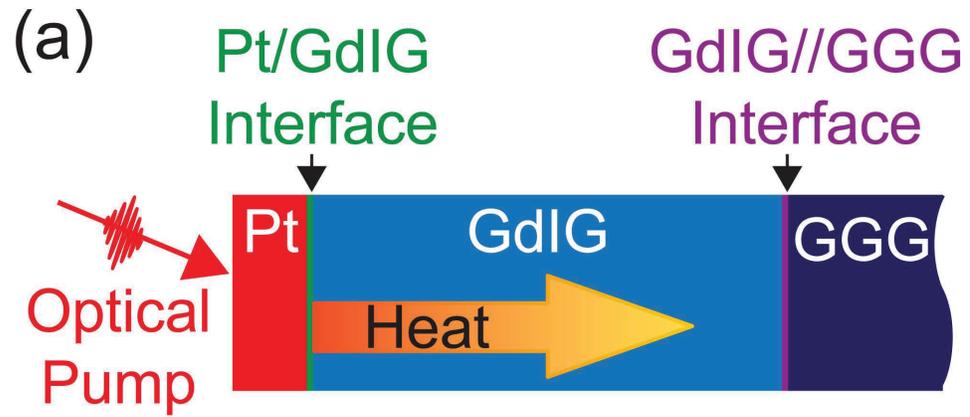

**(a)**

Pt/GdIG
Interface

GdIG//GGG
Interface

Optical
Pump

Pt    GdIG    GGG

Heat

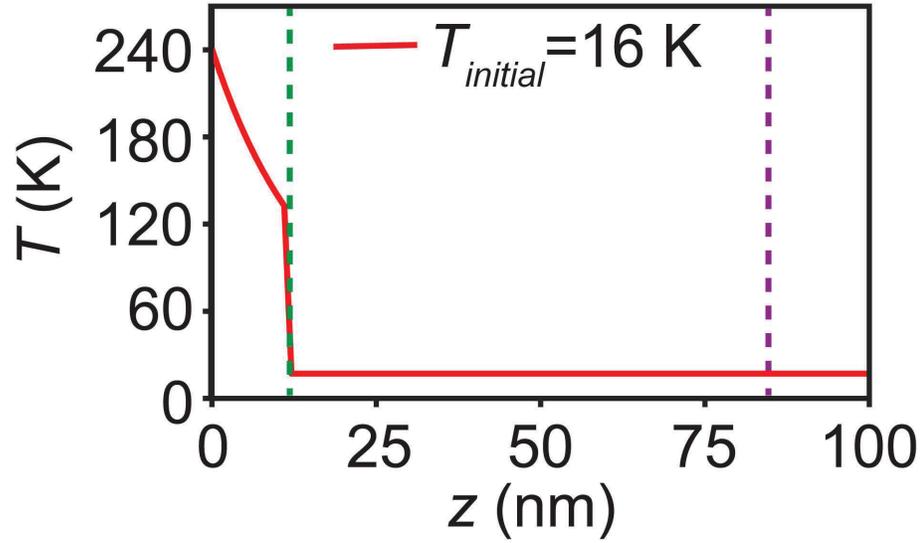

**(b)** $t$=0

$T_{initial}$=16 K

$T$ (K)

$z$ (nm)

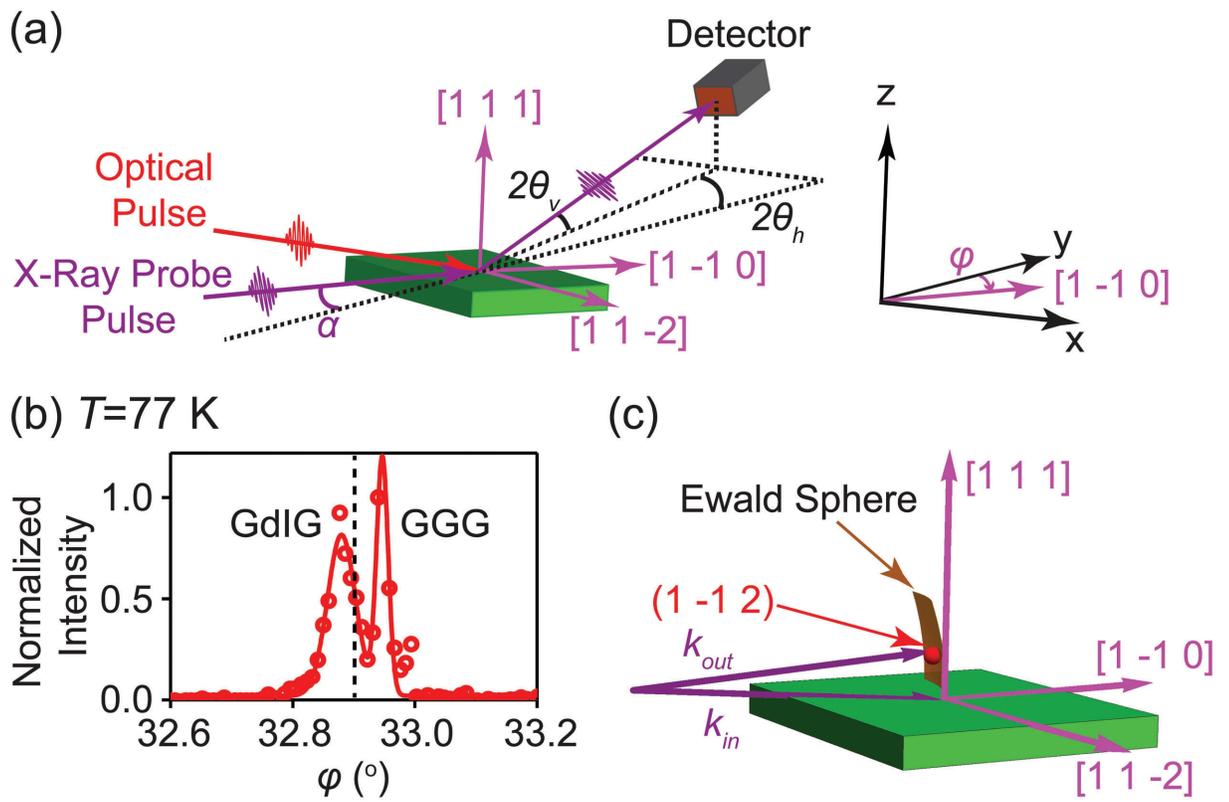

(a)

(b) *T*=77 K

(c)

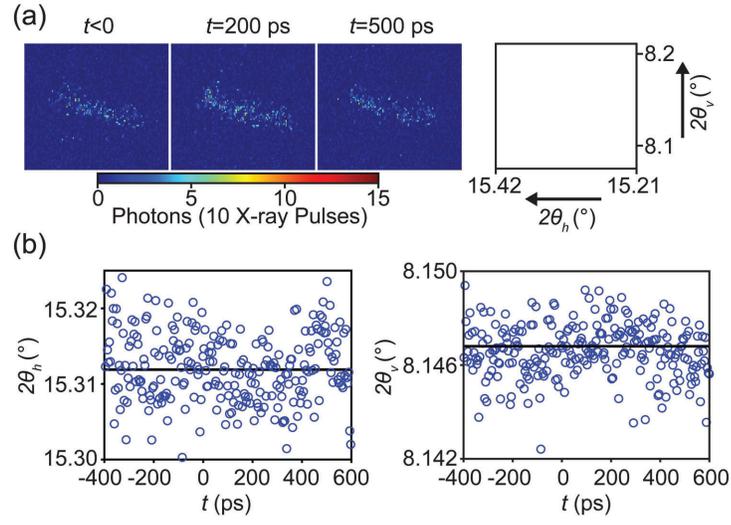

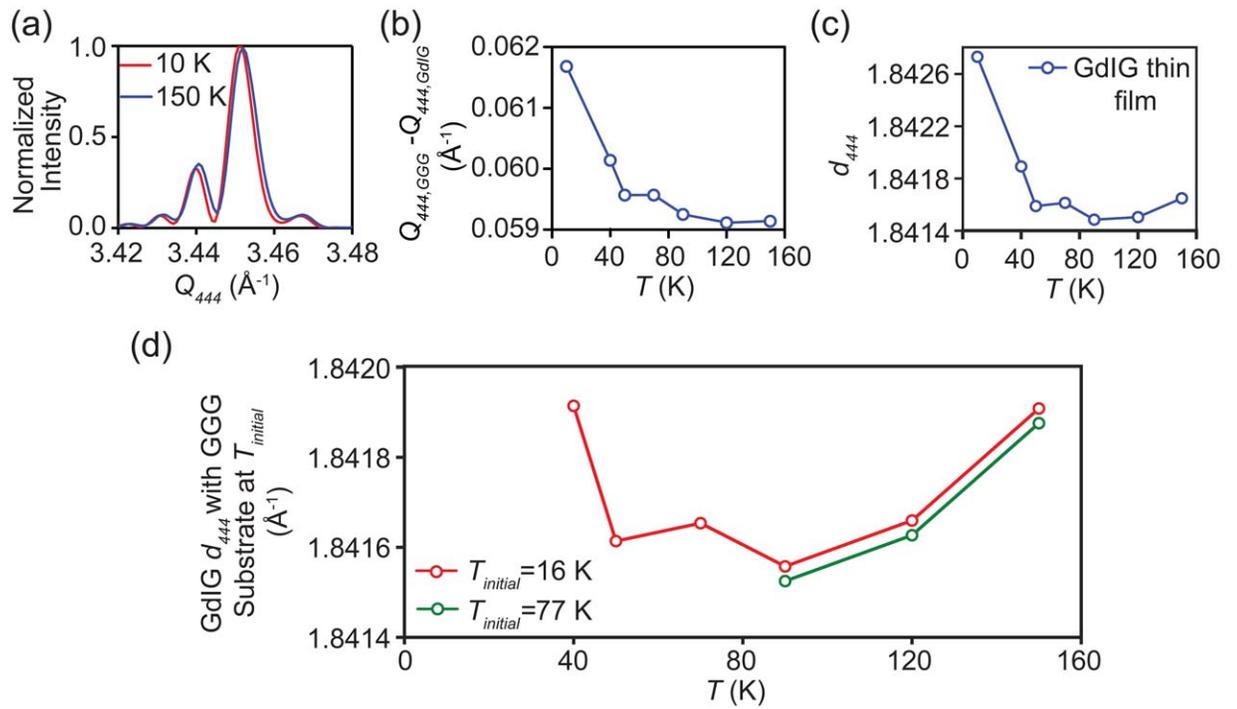

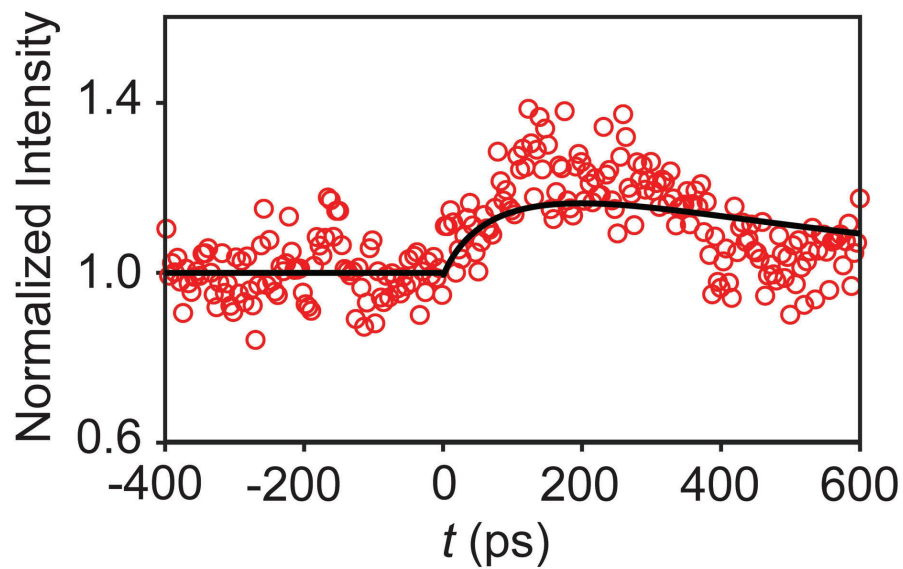

(a) $T_{initial}$=16 K

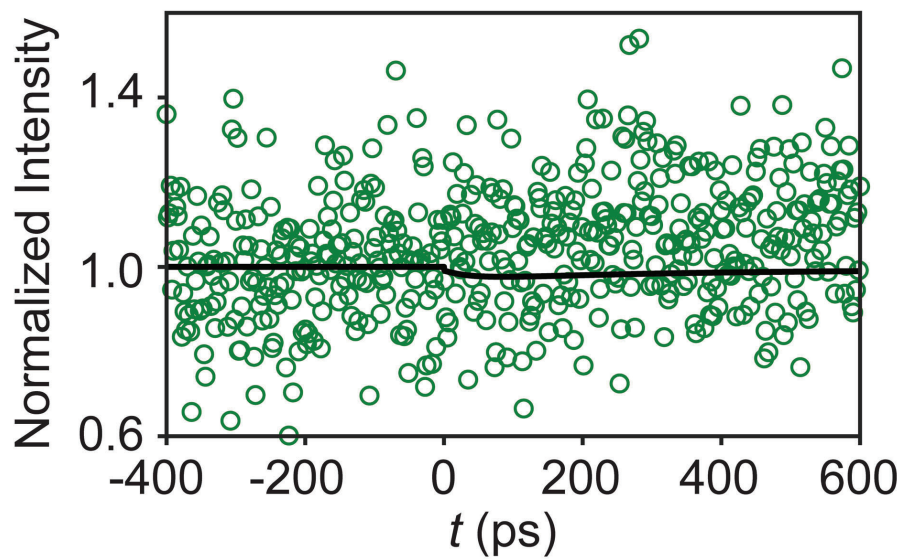

(b) $T_{initial}$=77 K

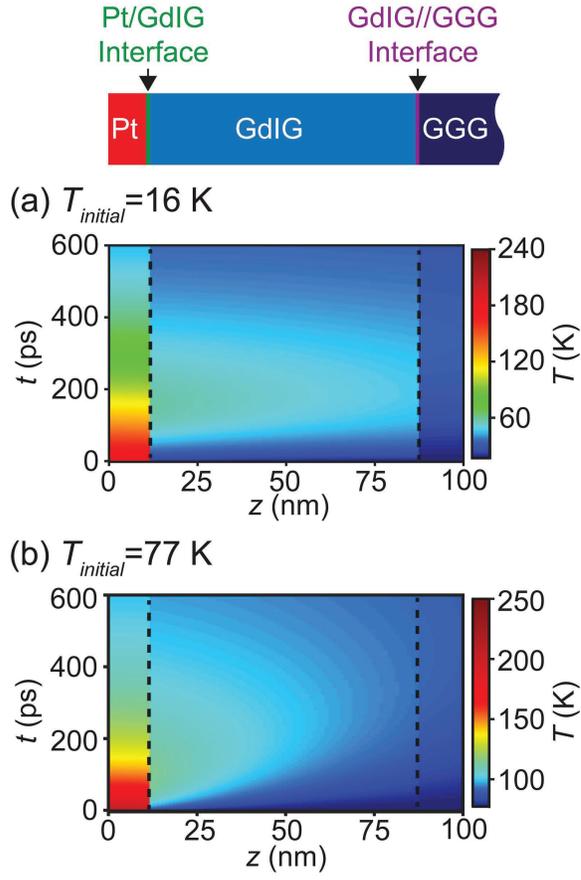

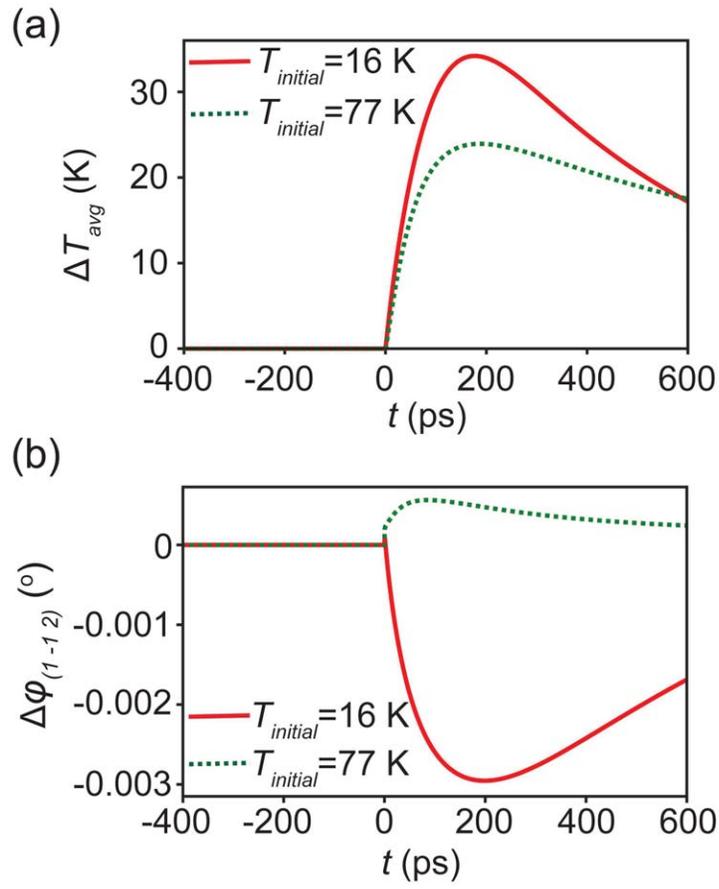